\documentclass[twocolumn,amsmath,amssymb,aps,prd,nofootinbib]{revtex4}

\usepackage{graphicx}
\usepackage{color} 

\begin{document}

\title{Late-time evolution of the Universe within a two-scalar-field cosmological model}

\author{Paulo M. S\'a}

\email{pmsa@ualg.pt}

\affiliation{Departamento de F\'{\i}sica, Faculdade de Ci\^encias e
Tecnologia, Universidade do Algarve, Campus de Gambelas, 8005-139 Faro,
Portugal}

\begin{abstract}
We investigate the late-time evolution of the Universe within a cosmological
model in which dark matter and dark energy are identified with two interacting
scalar fields. Using methods of qualitative analysis of dynamical systems, we
identify all cosmological solutions of this model. We show that viable
solutions---in the sense that they correspond to a cosmic evolution in which a
long enough matter-dominated era is followed by a current era of accelerated
expansion---can be found in several regions of the parameter space. These
solutions can be divided into two categories, namely, solutions that evolve to
a state of everlasting accelerated expansion, in which the energy density of
the dark-matter field rapidly approaches zero and the evolution becomes
entirely dominated by the dark-energy field, and solutions in which the stage
of accelerated expansion is temporary and the ratio between the energy
densities of dark energy and dark matter tends, asymptotically, to a constant
nonzero value.
\end{abstract}

\maketitle

\section{Introduction\label{Sect-introduction}}

One of the most striking developments of modern cosmology was the discovery
that the Universe is presently undergoing a period of accelerated expansion
\cite{riess-1998,perlmutter-1999}, dri\-ven by a still unknown form of energy,
called dark energy, which accounts for about 69\% of the total energy density
of the Universe \cite{Planck-parameters-2015}.

The simplest candidate for dark energy is the cosmological constant, whose
energy density remains unchanged throughout the evolution of the Universe.
Although consistent with current observational data, the cosmological constant
is unsatisfactory from the theoretical point of view, since its energy scale
estimated at both the classical and quantum levels strongly deviates from the
value required by cosmic observations \cite{weinberg-1989,martin-2012}.

An alternative and appealing approach is to consider the role of dark energy
to be played, not by a cosmological constant, but rather by a dynamical scalar
field, whose potential energy starts to dominate the evolution of the Universe
at later times, giving rise to a period of cosmic acceleration, in a way
similar to primordial inflation (for a review on dynamical dark energy see
Ref.~\cite{copeland-2006}).

A natural extension of this dynamical approach is to consider that dark
matter---whose physical nature, after decades of intense experimental efforts,
still remains unknown \cite{bertone-2018}---can also be identified with a
scalar field. This field could then, under certain circumstances, become the
dominant component of the Universe, giving rise to a matter-dominated era of
evolution needed for structure formation.

Identifying both dark energy and dark matter with scalar fields opens the
possibility to unify two seemingly disparate phenomena under the same
theoretical framework. This unifying effort can be taken even further if one
of the scalar fields also plays the role of the inflaton in the early Universe
(for such triple unifications see, for instance,
Refs.~\cite{capozziello-2006,liddle-2008,bose-2009,henriques-2009,santiago-2011,
odintsov-2019,lima-2019,sa-2020b,arbey-2021,oikonomou-2021}).

In a recent article \cite{sa-2020b}, a unified description of inflation, dark
energy, and dark matter was proposed within a two-scalar-field cosmological
model given by the action\footnote{We adopt in this article the natural system
of units and use the notation $\kappa\equiv \sqrt{8\pi G}=
\sqrt{8\pi}/m_\texttt{P}$, where $G$ is the gravitational constant and
$m_\texttt{P} = 1.22 \times 10^{19}\, {\rm GeV}$ is the Planck mass.}
\begin{align}
 S = {} & \int d^4x \sqrt{-g} \bigg[
  \frac{R}{2\kappa^2} - \frac12 (\nabla \phi)^2
    \nonumber
  \\
  & - \frac12 e^{-\alpha\kappa\phi} (\nabla \xi)^2
  - e^{-\beta\kappa\phi} V(\xi) \bigg],
 \label{action 2SF}
\end{align}
where $\alpha$ and $\beta$ are dimensionless parameters. Such an action arises
in a great variety of gravity theories, like the Jordan-Brans-Dicke theory,
Kaluza-Klein theories, $f(R)$-gravity, and string theories (see
Refs.~\cite{berkin-1991,starobinsky-2001,elizalde-2004} for a derivation of
the above action within these theories), as well as the hybrid metric-Palatini
theory \cite{harko-2012,tamanini-2013}. According to the proposed unification
scenario, for an appropriate choice of the potential $V(\xi)$ (see below),
inflation, assumed to be of the warm type, is driven by the scalar field
$\xi$, which shortly after the end of the inflationary period decouples from
radiation and starts behaving like a cold-dark-matter fluid; after a
radiation-dominated era, which encompasses the primordial nucleosynthesis
period, the dark-matter fluid, together with ordinary baryonic matter, gives
rise to a matter-dominated era, long enough to allow for structure formation;
finally, at recent times, the second scalar field $\phi$ emerges as the
dominant component of the Universe, giving rise to an era of accelerated
expansion. Resorting to numerical simulations, it was shown in
Ref.~\cite{sa-2020b} that, for certain values of the parameters $\alpha$ and
$\beta$, the picture emerging in this unified description of inflation, dark
energy, and dark matter is consistent with the standard cosmological model.

In the present article, we further investigate the two-scalar-field
cosmological model given by action~(\ref{action 2SF}), using the powerful
methods of qualitative analysis of dynamical systems. Our goal is to perform a
thorough investigation of the dynamical system arising in our model, covering
the entire parameter space $(\alpha,\beta)$, in order to identify all
solutions that reproduce the later stages of evolution of the Universe,
namely, the dark-matter- and dark-energy-dominated eras.

This article is organized as follows. In Sec.~\ref{Sect-2SFCM} we pre\-sent
the evolution equations for the two-scalar-field cosmological model.
Section~\ref{Sect-DynSys} and Appendix~\ref{Sect-appendix_1} are devoted to
the stability analysis of the critical points of the dynamical system arising
in our cosmological model. In Sec.~\ref{Sect-CosmSol} we interpret the results
of this analysis and identify all solutions that correspond to viable
cosmological scenarios. Finally, in Sec.~\ref{Sect-conclusions}, we present
our conclusions.

\section{Two-scalar-field \protect\\ cosmological model \label{Sect-2SFCM}}

Let us consider a cosmological model described by action~(\ref{action 2SF}).
Varying this action with respect to $g_{\mu\nu}$, $\phi$ and $\xi$ and
assuming a flat Friedmann--Lema\^{i}tre--Robertson\--Walker
metric\footnote{Because the curvature density parameter $\Omega_k$ is
constrained by current cosmological measurements to be very small
\cite{Planck-parameters-2015}, we can assume a spatially flat metric without
much loss of generality.}, we obtain the Einstein equations for the scale
factor $a(t)$
\begin{align}
& \bigg( \frac{\dot{a}}{a} \bigg)^2 = \frac{\kappa^2}{3} \bigg(
 \frac{\dot{\phi}^2}{2} + \frac{\dot{\xi}^2}{2} e^{-\alpha\kappa\phi}
 + V e^{-\beta\kappa\phi} \bigg), \label{dot-a-1}
   \\
& \hspace{2.7mm}\frac{\ddot{a}}{a} =  - \frac{\kappa^2}{3} \bigg( \dot{\phi}^2 +
 \dot{\xi}^2 e^{-\alpha\kappa\phi} - V e^{-\beta\kappa\phi} \bigg),
 \label{ddot-a-1}
\end{align}
and the equations of motion for the scalar fields $\xi(t)$ and $\phi(t)$
\begin{align}
 & \ddot{\xi} + 3 \frac{\dot{a}}{a} \dot{\xi} - \alpha\kappa \dot{\phi}\dot{\xi}
 + \frac{\partial V}{\partial \xi} e^{(\alpha-\beta)\kappa\phi}
 = 0, \label{ddot-xi-1}
  \\
 & \ddot{\phi} + 3 \frac{\dot{a}}{a} \dot{\phi} + \frac{\alpha\kappa}{2}
 \dot{\xi}^2 e^{-\alpha\kappa\phi}
 -\beta \kappa V e^{-\beta \kappa \phi}
 = 0, \label{ddot-phi-1}
\end{align}
where an overdot denotes a derivative with respect to time $t$. These two last
equations differ from the usual ones in that they contain an extra term
arising due to the presence in action~(\ref{action 2SF}) of a nonstandard
kinetic term for the scalar field $\xi$; the usual equations are recovered for
$\alpha=0$.

Choosing the potential $V(\xi)$ to be of the form\footnote{Within scalar-field
models for the interaction of dark energy and dark matter, several potentials
have been considered in the literature; for a discussion of a generic
potential of the form $V(\phi,\xi)=e^{-\lambda\phi}P(\phi,\xi)$, where
$P(\phi,\xi)$ is a polynomial, see Ref.~\cite{bertolami-2012}.}
\begin{equation}
 V(\xi)  = V_a + \frac12 m^2 \xi^2,
  \label{potential xi}
\end{equation}
where $V_a$ and $m$ are constants, and defining the energy density and
pressure of the scalar fields as
\begin{align}
 \rho_\xi &= \frac{\dot{\xi}^2}{2}
 e^{-\alpha\kappa\phi}
 + \frac12 m^2 e^{-\beta\kappa\phi} \xi^2,
  \label{rho xi E1}
\\
 \rho_\phi &= \frac{\dot{\phi}^2}{2}
 + V_a e^{-\beta\kappa\phi},
  \label{rho phi E1}
  \end{align}
and
\begin{align}
 p_\xi &= \frac{\dot{\xi}^2}{2}
 e^{-\alpha\kappa\phi}
 - \frac12 m^2 e^{-\beta\kappa\phi} \xi^2,
    \label{p xi E1}
 \\
  p_\phi &= \frac{\dot{\phi}^2}{2}
 - V_a e^{-\beta\kappa\phi},
  \label{p phi E1}
\end{align}
we can write Eq.~(\ref{ddot-xi-1}) for the scalar field $\xi$ as
\begin{equation}
  \dot{\rho}_\xi
  + 3 H (\rho_\xi+p_\xi)
   =\frac{\kappa}{2} \Big( \alpha \dot{\xi}^2  e^{-\alpha\kappa\phi}
  -\beta m^2 \xi^2  e^{-\beta\kappa\phi} \Big) \dot{\phi},
    \label{ddot-xi-2}
\end{equation}
where $H=\dot{a}/a$ is the Hubble parameter.

Let us now assume that the scalar field $\xi$ oscillates rapidly around the
minimum of its potential, thus behaving like a nonrelativistic dark-matter
fluid with an equation of state $\langle p_\xi \rangle=0$, where the brackets
$\langle ... \rangle$ denote the average over an oscillation period. Then,
averaging over an oscillation and taking into account that $\big<p_\xi\big>=0$
implies $\big< \xi^2 \big>=\rho_\xi m^{-2} e^{\beta\kappa\phi}$ and $\big<
\dot{\xi}^2 \big> = \rho_\xi e^{\alpha\kappa\phi}$, the evolution equations
become
\begin{align}
 & \dot{\rho}_\xi + 3H \rho_\xi
  = \frac{\kappa}{2} (\alpha-\beta) \rho_\xi \dot{\phi},
    \label{dot-rhoxi}
\\
 & \ddot{\phi}+3H\dot{\phi}-\beta\kappa V_a e^{-\beta\kappa\phi}
  = -\frac{\kappa}{2} (\alpha-\beta) \rho_\xi,
    \label{ddot-phi-3}
\\
 & \dot{H}=-\frac{\kappa^2}{2} \left( \dot{\phi}^2 + \rho_\xi \right),
    \label{dot-H}
\end{align}
subject to the Friedmann constraint
\begin{equation}
  H^2 = \frac{\kappa^2}{3} \bigg( \frac{\dot{\phi}^2}{2}
 + V_a e^{-\beta\kappa\phi} + \rho_\xi \bigg).
    \label{friedmann}
\end{equation}

In previous work \cite{sa-2020b}, we have shown that this two-scalar-field
cosmological model allows for a unified description of inflation, dark matter,
and dark energy, in which the scalar field $\xi$ plays the roles of both
inflaton and dark matter, while the scalar field $\phi$ plays the role of dark
energy\footnote{In the specific case $\alpha=\sqrt6/3$ and arbitrary $\beta$,
corresponding to a generalized hybrid metric-Palatini theory of gravity, a
unified description of dark matter and dark energy---but not inflation---was
proposed in Ref.~\cite{sa-2020a}. For other cosmological solutions and the
weak-field limit of this theory, see Refs.~\cite{rosa_2020,bombacigno-2019}.}.
There, the solution of Eq.~(\ref{dot-rhoxi}),
\begin{equation}
 \rho_\xi = \rho_{\xi,0} \left( \frac{a_0}{a} \right)^3
 e^{\frac{\kappa}{2}(\alpha-\beta)(\phi-\phi_0)},
 \label{solution-rho_xi}
\end{equation}
where the subscript 0 denotes present-time quantities, was inserted into
Eqs.~(\ref{ddot-phi-3})--(\ref{friedmann}) and the resulting system was solved
numerically for specific values of $\alpha$ and $\beta$.

In the present article, instead of numerical methods, we use methods of
qualitative analysis of dynamical systems to investigate the solutions of
Eqs.~(\ref{dot-rhoxi})--(\ref{friedmann}), covering now the entire parameter
space ($\alpha,\beta$).

To conclude this section, let us point out that the two-scalar-field
cosmological model under consideration admits a direct transfer of energy
between dark energy and dark matter, mediated by the term
\begin{equation}
Q =\frac{\kappa}{2} (\alpha-\beta) \rho_\xi \dot{\phi},
 \label{Q-interaction}
\end{equation}
as results from Eqs.~(\ref{dot-rhoxi}) and (\ref{ddot-phi-3}).

Cosmological models with an interaction term $Q\propto \rho \dot{\phi}$, where
$\phi$ is a scalar field with an exponential potential and $\rho$ is the
energy density of a perfect fluid with an equation of state $p=w\rho$, have
been investigated by several authors
\cite{amendola-1999,holden-2000,billyard-2000,amendola-2000,
tocchini-valentini-2002,gumjudpai-2005,boehmer-2008,tzanni-2014,singh-2016,
bernardi-2017} (for other interaction terms considered in dark-matter and
dark-energy interaction models, see the review
articles~\cite{bolotin-2015,wang-2016}).
Although the present paper is focused on the background dynamics of the
cosmological model given by Eqs.~(\ref{dot-rhoxi})--(\ref{friedmann}) and on its capability to reproduce the late-time evolution of the Universe, it is worth emphasizing that such interaction models have also been studied at the perturbative level (linear and nonlinear), with the conclusion that they are compatible with observations of the microwave background radiation and cosmic structure formation \cite{amendola-2000,amendola-2003,maccio-2004,mainini-2006,brookfield-2008,
pettorino-2008,xia-2009,baldi-2010,tarrant-2012,pettorino-2013,barros-2019}.

What is rather interesting in the interaction term Q given by
Eq.~(\ref{Q-interaction}) is that it vanishes for $\alpha=\beta$. This means
that the transfer of energy between two scalar fields directly coupled via an
exponential potential can be canceled due to the presence of a nonstandard
kinetic term. In this case, the energy density of dark matter $\rho_\xi$
evolves as $a^{-3}$ [see Eq.~(\ref{solution-rho_xi})], i.e., exactly as
ordinary baryonic matter, while dark energy evolves subject only to the
potential $V_a e^{-\kappa\beta\phi}$.
For $\alpha \neq \beta$ there is a direct energy exchange between the two
scalar fields, implying that dark matter, although pressureless, does not
scale simply as $a^{-3}$; it depends also on the dark-energy field $\phi$.
Such dependence has consequences on the cosmic evolution,
more specifically, the energy density of dark energy becomes a non-negligible
fraction of the total energy density throughout the matter-dominated era and
the transition from radiation to matter domination occurs earlier in the cosmic
history. As shown in Ref.~\cite{sa-2020b}, resorting to numerical simulations, to avoid a
conflict with primordial nucleosynthesis, the condition $|\alpha-\beta|\lesssim1$
must be imposed on the parameters of the model (we will return to this
issue in Sec.~\ref{Sect-CosmSol}).

\section{Dynamical-system analysis \label{Sect-DynSys}}

Let us now turn to the analysis of the system of differential
equations~(\ref{dot-rhoxi})--(\ref{friedmann}) using methods of qualitative
analysis of dynamical systems (for a recent review on dynamical systems
applied to cosmology and, in particular, to dark-energy models, see
Ref.~\cite{bahamonde-2018}).

Following Ref.~\cite{copeland-1998}, we introduce the dimensionless variables,
\begin{equation}
  x = \frac{\kappa}{\sqrt6 H} \dot{\phi}
  \quad \mbox{and} \quad
  y = \frac{\kappa}{\sqrt3 H} \sqrt{V_a e^{-\beta\kappa\phi}},
\end{equation}
as well as a new time variable $\tau$, defined as
\begin{equation}
  \tau = \ln a, \qquad \frac{d\tau}{dt}=H.
\end{equation}

In these new variables, the system of
equations~(\ref{dot-rhoxi})--(\ref{friedmann}) can be written as
\begin{subequations}
 \label{DS}
\begin{align}
 x^\prime  = &-3x + \frac{\sqrt{6}}{2} \beta y^2 + \frac32 x(1+x^2-y^2)
  \nonumber
  \label{x}
\\
     & -\frac{\sqrt{6}}{4} (\alpha-\beta) (1-x^2-y^2),
\\
 y^\prime = &-\frac{\sqrt{6}}{2} \beta x y + \frac32 y(1+x^2-y^2),
 \label{y}
\end{align}
\end{subequations}
where the prime denotes a derivative with respect to the logarithmic time
$\tau$.
Notice that the Friedmann constraint, given by
\begin{equation}
   x^2+y^2+\Omega_\xi=1, \quad \Omega_\xi\equiv \rho_\xi \frac{\kappa^2}{3H^2},
     \label{friedmann xy}
\end{equation}
was used in Eqs.~(\ref{DS})
to eliminate the explicit dependence on the variable $\rho_\xi$,
thus reducing the dynamical system to just two dimensions, a circumstance
that considerably simplifies the analysis.

Because the density parameter $\Omega_\xi$ is, by definition, non-negative
and we are assuming a flat universe, it follows from the Friedmann
constraint~(\ref{friedmann xy}) that the variables $x$ and $y$ should
satisfy the condition $x^2+y^2 \leq 1$,
i.e., the physically relevant orbits of the dynamical system~(\ref{DS})
are confined to the unit circle.
Furthermore, since we are interested in expanding cosmologies, the analysis
should be restricted to the upper semicircle, for which $y\geq0$.
In summary, the phase space of the dynamical system~(\ref{DS}) is the upper
half of the unit circle centered at the origin.

The dynamical system~(\ref{DS}) contains two dimensionless constants $\alpha$
and $\beta$, which parameterize the nonstandard kinetic term of the scalar
field $\xi$ and the interaction potential between the scalar fields $\phi$ and
$\xi$, respectively [see action~(\ref{action 2SF})]. Without any loss of
generality, we can assume that $\alpha$ is non-negative\footnote{Indeed, since
the dynamical system~(\ref{DS}) is invariant under the transformation
$x\rightarrow -x$, $\alpha\rightarrow -\alpha$, and $\beta\rightarrow -\beta$,
solutions for negative values of $\alpha$ ($\beta$) can be obtained
straightforwardly from solutions for positive values of $\alpha$ ($\beta$),
provided a reflection over $x$ is performed, as well as a change of sign of
the parameter  $\beta$ ($\alpha$).}. Furthermore, the case
$\alpha=0$, corresponding to an action with a standard kinetic term for the
scalar field $\xi$, has been extensively studied in the literature
\cite{bahamonde-2018} and, therefore, will not be considered in this article.
In what concerns $\beta$, we allow it to take any value, including the value
zero, for which the direct coupling in the potential between the two scalar
fields vanishes. In summary, the parameters $\alpha$ and $\beta$ span the open
half-plane $\alpha>0$.

\begin{figure*}
 \includegraphics{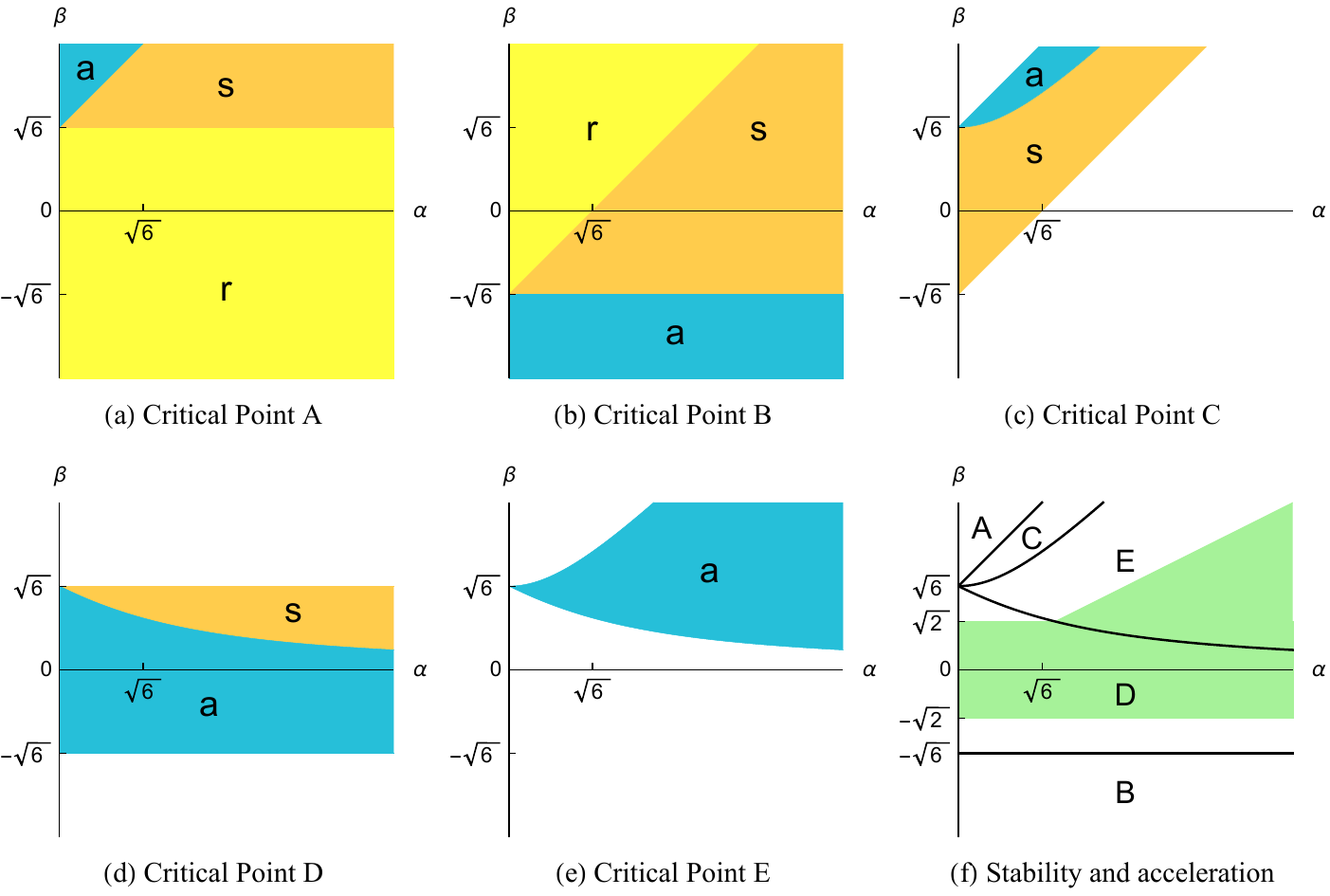}
 \caption{\label{Fig:critical points} Panels (a) to (e) show the regions of
existence and stability of the critical points of the dynamical
system~(\ref{DS}) in the parameter space $(\alpha,\beta)$---the open
half-plane $\alpha>0$. In white regions, critical points do not exist. In blue
(dark shaded) regions they exist and are stable (attractors), while in orange
(medium shaded) and yellow (lightly shaded) regions they exist, but are
unstable (saddles and repellers, respectively). The regions in which the
critical points are attractors, saddles, and repellers, are also denoted by
the letters ``a", ``s", and ``r", respectively. Panel (f) shows the parameter
space divided into five regions, in each of which one and only one of the
critical points A, B, C, D, and E is an attractor. The part of the parameter
space in which critical point D (or critical point E) is an attractor and,
simultaneously, corresponds to a state of accelerated expansion is highlighted
in green color (shaded).}
\end{figure*}

Depending on the values of $\alpha$ and $\beta$, the dynamical
system~(\ref{DS}) has up to five critical points. Table~\ref{Table:properties
CP} and Fig.~\ref{Fig:critical points} summarize, in both analytical and
graphical form, the conditions for their existence and stability, as well as
the conditions for the existence of accelerated solutions.

\begin{table*}
\begin{tabular}{ccccccc}
\hline\hline
  Point
  & \quad $x$
  & \quad $y$
  & \quad Existence
  & \quad Stability
  & \quad Acceleration \\ \hline \noalign{\vskip 1mm}
A & \quad $1$
  & \quad $0$
  & \quad $\forall\alpha,\beta$
  & \quad $\beta \geq \alpha + \sqrt6$
  & \quad never \\ \noalign{\vskip 1mm}
B & \quad $-1$
  & \quad $0$
  & \quad $\forall\alpha,\beta$
  & \quad $\beta \leq -\sqrt6$
  & \quad never \\ \noalign{\vskip 1mm}
C & \quad $\frac{\beta-\alpha}{\sqrt6}$
  & \quad $0$
  & \quad $|\alpha-\beta|\leq \sqrt{6}$
  & \quad $\sqrt{\alpha^2 + 6} \leq \beta \leq \alpha +\sqrt6$
  & \quad never \\ \noalign{\vskip 1mm}
D & \quad $\frac{\beta}{\sqrt6}$
  & \quad $\sqrt{1-\frac{\beta^2}{6}}$
  & \quad $|\beta|\leq\sqrt6$
  & \quad $-\sqrt6 \leq \beta \leq \frac{-\alpha+\sqrt{\alpha^2+24}}{2}$
  & \quad $|\beta|<\sqrt2$ \\ \noalign{\vskip 1mm}
E & \quad $\frac{\sqrt6}{\alpha+\beta}$
  & \quad $\frac{\sqrt{6+\alpha^2-\beta^2}}{\alpha+\beta}$
  & \quad $\frac{-\alpha+\sqrt{\alpha^2+24}}{2} \leq \beta \leq \sqrt{\alpha^2+6}$
  & \quad $\frac{-\alpha+\sqrt{\alpha^2+24}}{2}\leq \beta \leq \sqrt{\alpha^2 + 6}$
  & \quad $\beta <\frac{\alpha}{2}$ \\ \noalign{\vskip 1mm}
\hline\hline
\end{tabular}
 \caption{\label{Table:properties CP} Properties of the critical points of the
 dynamical system~(\ref{DS}): existence, stability, and acceleration.}
 \end{table*}

For most values of $\alpha$ and $\beta$, the stability of the
critical points can be assessed simply by using the linear theory, since, for
those values, both eigenvalues of the Jacobian matrix of the dynamical
system~(\ref{DS}) have a nonzero real part. However, for certain values of
$\alpha$ and $\beta$, namely, $\beta=\alpha\pm\sqrt{6}$,
$\beta=\sqrt{\alpha^2+6}$, $\beta=\pm\sqrt{6}$, and
$\beta=(-\alpha+\sqrt{\alpha^2+24})/2$, corresponding to the lines delimiting
the different regions of the plots of Fig.~\ref{Fig:critical points}, one
of the eigenvalues becomes zero, forcing us to go beyond the linear theory and
use other methods to study the stability properties of the critical points,
such as center manifold theory and Lyapunov's method. In
Appendix~\ref{Sect-appendix_1} we present a full analysis of the stability of
the five critical points for all values of $\alpha$ and $\beta$ belonging to
the parameter space.

Note that each of the critical points is stable for specific values of
$\alpha$ and $\beta$ and that to each point of the parameter space
$(\alpha,\beta)$ corresponds one and only one stable critical point (see
Fig.~\ref{Fig:critical points}).

\begin{figure}
 \includegraphics[width=7.9cm]{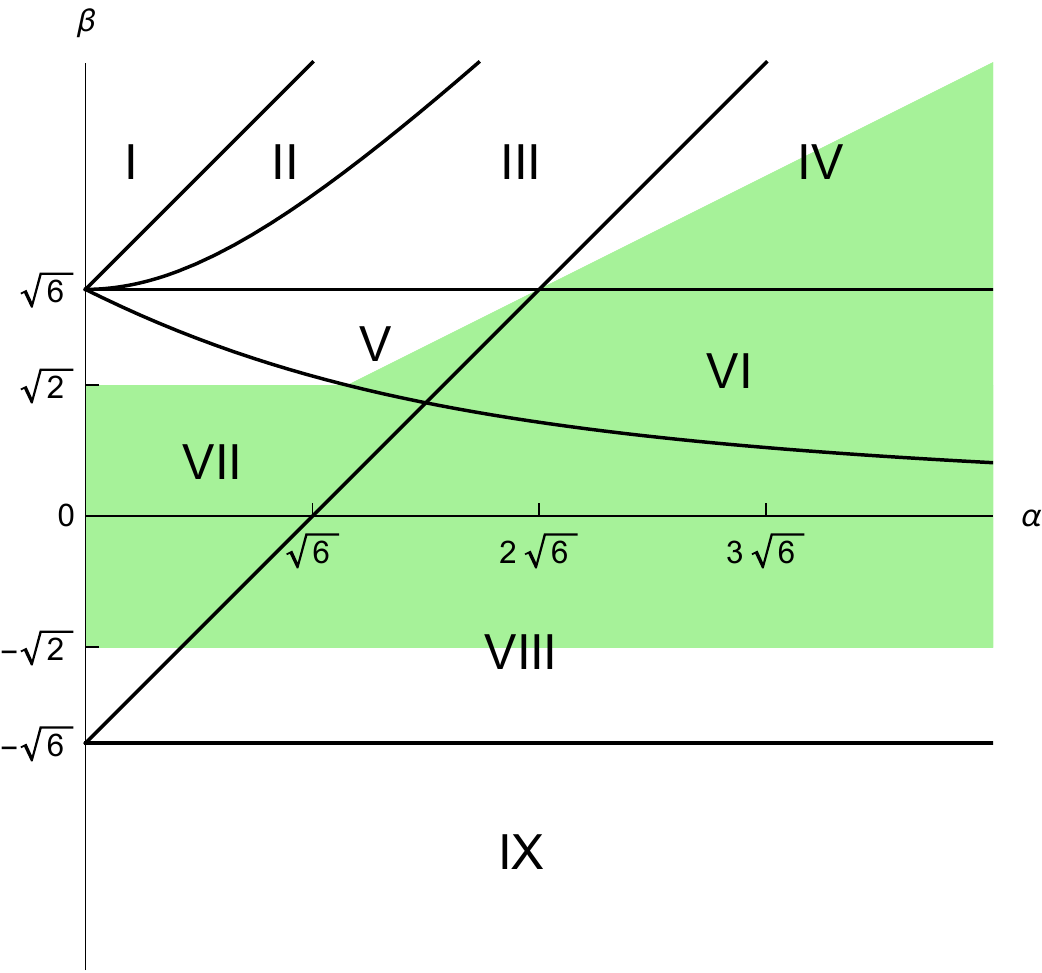}
 \caption{ \label{Fig:9 regions} The existence and stability conditions of the
critical points allow for a division of the parameter space $(\alpha,\beta)$
into nine regions, labeled with Roman numerals, each of which corresponds to a
qualitatively different phase portrait of the dynamical system~(\ref{DS}). In
green color (shaded) it is indicated the regions or part of them in which
critical point D (or critical point E) is an attractor and, simultaneously,
corresponds to a state of accelerated expansion.}
\end{figure}

\begin{figure*}
 \includegraphics{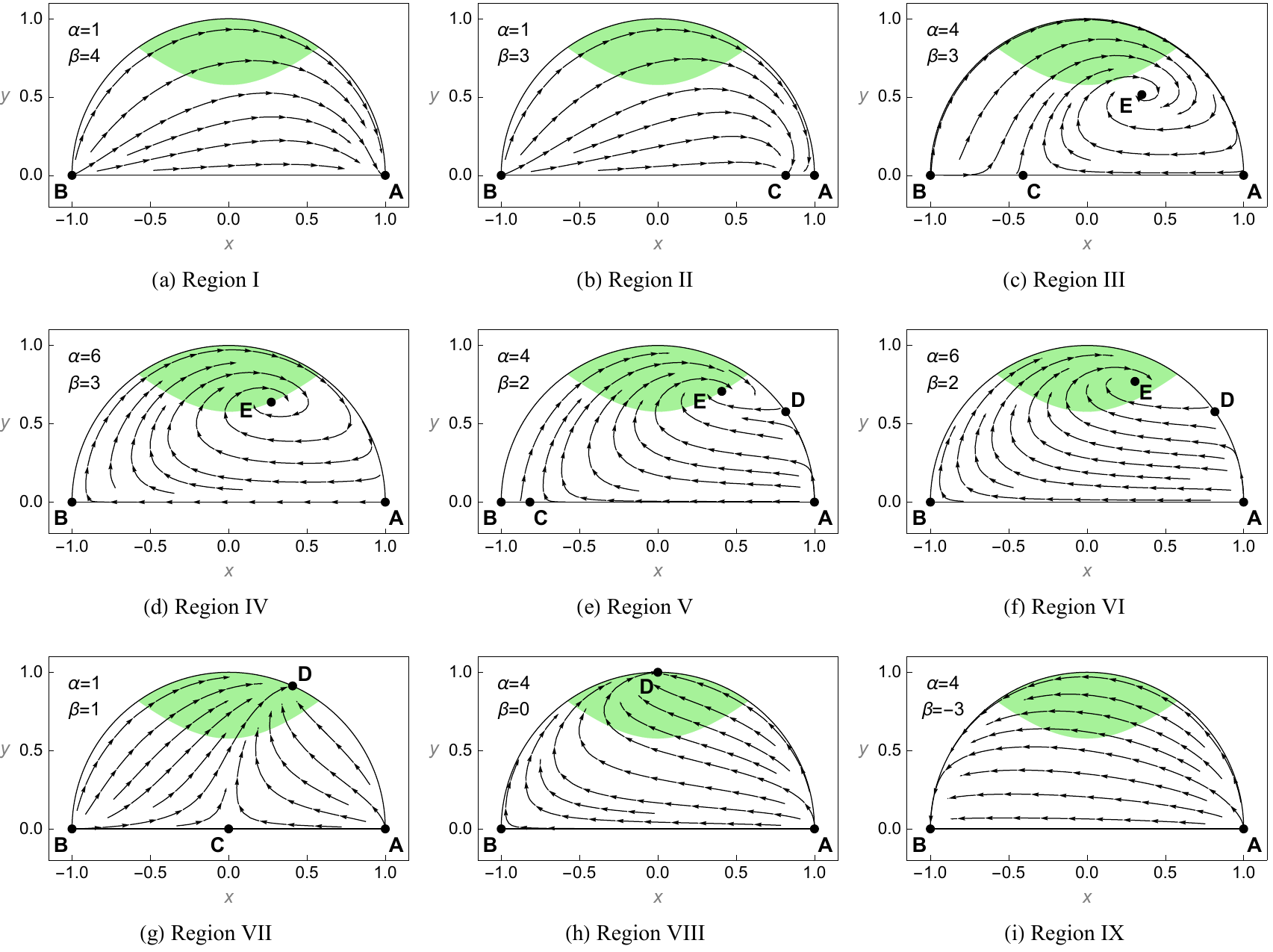}
 \caption{\label{Fig:phase portraits} Phase portraits of the dynamical
system~(\ref{DS}) for different values of the parameters $\alpha$ and $\beta$.
Each of these phase portraits corresponds to one of the nine regions of the
parameter space shown in Fig.~\ref{Fig:9 regions} (see also Table~\ref{Table:9
regions}). In the green (shaded) region of phase space the expansion of the
Universe is accelerated.}
\end{figure*}

The existence and stability conditions for the critical points allow for a
division of the parameter space $(\alpha,\beta)$ into nine regions (see
Fig.~\ref{Fig:9 regions} and Table~\ref{Table:9 regions}), each of which
corresponds to a qualitatively different phase portrait. Nine phase portraits,
representative of the behavior of the dynamical system~(\ref{DS}) in each of
these regions, are shown in Fig.~\ref{Fig:phase portraits}.

\begin{table}
\begin{tabular}{cccccccccc}
\hline\hline
Point
  & \quad I
  & \quad II
  & \quad III
  & \quad IV
  & \quad V
  & \quad VI
  & \quad VII
  & \quad VIII
  & \quad IX
  \\ \hline \noalign{\vskip 1mm}
A & \quad a
  & \quad s
  & \quad s
  & \quad s
  & \quad r
  & \quad r
  & \quad r
  & \quad r
  & \quad r
  \\ \noalign{\vskip 1mm}
B & \quad r
  & \quad r
  & \quad r
  & \quad s
  & \quad r
  & \quad s
  & \quad r
  & \quad s
  & \quad a
  \\ \noalign{\vskip 1mm}
C & \quad
  & \quad a
  & \quad s
  & \quad
  & \quad s
  & \quad
  & \quad s
  & \quad
  & \quad
  \\ \noalign{\vskip 1mm}
D & \quad
  & \quad
  & \quad
  & \quad
  & \quad s
  & \quad s
  & \quad a
  & \quad a
  & \quad
  \\ \noalign{\vskip 1mm}
E & \quad
  & \quad
  & \quad a
  & \quad a
  & \quad a
  & \quad a
  & \quad
  & \quad
  & \quad
  \\ \noalign{\vskip 1mm}
  \hline\hline
\end{tabular}
 \caption{\label{Table:9 regions} Division of the parameter space
$(\alpha,\beta)$ into nine regions, labeled with Roman numerals. For each such
region, the critical points are classified using the letters ``a", ``s", and
``r", denoting attractor, saddle,  and repeller, respectively. The absence of
a letter means that the corresponding  critical point does not exist in that
region.}
\end{table}

Since there is only one attracting critical point in each region, all orbits
of the corresponding phase portraits (except the heteroclinic ones connecting
the other critical points) asymptotically converge to this unique attractor,
which, therefore, represents the final stage of evolution of the Universe. The
phase portraits corresponding to regions II to VIII always contain at least
one saddle point (see Table~\ref{Table:9 regions}). The repellers are always
critical points A and/or B. The exception is the phase portrait of region IV,
which has no repellers at all; for $\tau\rightarrow -\infty$, all orbits
approach the heteroclinic one connecting critical points A and B through the
$x$-axis and the upper half of the circumference $x^2+y^2=1$. In regions V and
VII, the phase portraits have two repellers, critical points A and B. In this
case, the phase space is divided into two parts by the heteroclinic orbit
connecting points C, D, and E (region V) and points C and D (region VII);
orbits on the right part of the phase space originate at critical point A,
while on the left they have their origin at critical point~B.

Before proceeding to the physical interpretation of the results obtained in
this section, let us point out that, in the variables $x$ and $y$,
the density parameter for the scalar field
$\phi$, the effective equation-of-state parameter, and the deceleration
parameter, are given by
\begin{equation}
 \Omega_\phi \equiv \rho_\phi \frac{\kappa^2}{3H^2}  = x^2+y^2,
\end{equation}
\begin{equation}
 w_{\rm eff} \equiv \frac{p_\xi + p_\phi}{\rho_\xi + \rho_\phi} = x^2 - y^2,
\end{equation}
and
\begin{equation}
 q \equiv - \frac{H^\prime}{H} -1 = \frac12 \left( 1+3x^2 - 3y^2 \right),
  \label{q}
\end{equation}
respectively.

For certain values of $\alpha$ and $\beta$, critical points D and E correspond
to a state in which the deceleration parameter $q$ is negative. In the phase
space $(x,y)$, this region of negative $q$ lies above the curve
$y=\sqrt{x^2+1/3}$ (see Fig.~\ref{Fig:phase portraits}); orbits inside this
region correspond to a state of accelerated expansion of the Universe. The
regions of the parameter space $(\alpha,\beta)$ in which critical points D and
E are attractors and, simultaneously, correspond to a state of accelerated
expansion are shown in panel (f) of Fig.~\ref{Fig:critical points} and in
Fig.~\ref{Fig:9 regions}.

\section{Cosmological solutions \label{Sect-CosmSol}}

Let us now proceed to the physical interpretation of the results obtained in
Sec.~\ref{Sect-DynSys}.

Critical points A and B correspond to a state of the Universe in which the
total energy density $\rho=\rho_\xi+\rho_\phi$ is dominated by the kinetic
term of the scalar field $\phi$ ($\Omega_\phi=1$), which, therefore, behaves
as a stiff-matter fluid ($w_{\rm eff}=1$). The density parameter $\Omega_\xi$
vanishes, meaning that the influence of the scalar field $\xi$ on the dynamics
of the Universe is negligible in the vicinity of these critical points (see
Table~\ref{Table:properties CP-2}).

\begin{table}[t]
\begin{tabular}{cccc}
\hline\hline
  Point
  & \quad $\Omega_\phi$
  & \quad $\Omega_\xi$
  & \quad $w_{\rm eff}$ \\ \hline \noalign{\vskip 1mm}
A & \quad $1$
  & \quad $0$
  & \quad $1$ \\ \noalign{\vskip 1mm}
B & \quad $1$
  & \quad $0$
  & \quad $1$ \\ \noalign{\vskip 1mm}
C & \quad $\frac{(\alpha-\beta)^2}{6}$
  & \quad $1-\frac{(\alpha-\beta)^2}{6}$
  & \quad $\frac{(\alpha-\beta)^2}{6}$ \\ \noalign{\vskip 1mm}
D & \quad $1$
  & \quad $0$
  & \quad $-1+\frac{\beta^2}{3}$ \\ \noalign{\vskip 1mm}
E & \quad $\frac{12+\alpha^2-\beta^2}{(\alpha+\beta)^2}$
  & \quad $\frac{2(\alpha\beta+\beta^2-6)}{(\alpha+\beta)^2}$
  & \quad $\frac{-\alpha+\beta}{\alpha+\beta}$ \\ \noalign{\vskip 1mm}
    \hline\hline
\end{tabular}
 \caption{\label{Table:properties CP-2} Density parameters for dark energy
and dark matter and the effective equation-of-state parameter for the critical
points of the dynamical system~(\ref{DS}).}
\end{table}

At critical point C, the state of the Universe depends crucially on the values
of the parameters $\alpha$ and $\beta$. For $\alpha=\beta$, the dark-matter
field $\xi$ dominates the evolution of the Universe ($\Omega_\xi=1$). For
$\beta=\alpha\pm\sqrt6$, corresponding to the lines delimiting the region of
existence of this critical point, it is the kinetic term of the scalar field
$\phi$ that  dominates ($\Omega_\phi=1$), leading to a stiff-matter behavior.
For intermediate values of the difference $\alpha-\beta$, neither $\xi$ nor
$\phi$ entirely dominate the evolution of the Universe, and, consequently, the
effective equation-of-state parameter $w_{\rm eff}$ can take any value in the
range $0$ to $1$, depending on the relative preponderance of the scalar
fields. Of particular relevance is the case in which critical point C is a
saddle and $\xi$ is the dominant field (such behavior may occur in regions
III, V, and VII of the parameter space); it corresponds to a transitory period
of matter domination in the history of the Universe, needed for structure
formation. Although of lesser relevance, let us note that critical point C can
also be an attractor (in region II of the parameter space), in which case the
ratio $\Omega_\phi/\Omega_\xi$ between the density parameters of the scalar
fields becomes locked, asymptotically, at a constant value determined by
$\alpha$ and $\beta$ (see Table~\ref{Table:properties CP-2}).

At critical point D, which exists for $|\beta|\leq\sqrt6$, the evolution of
the Universe is dominated by the scalar field $\phi$ ($\Omega_\phi=1$). The
effective equation-of-state parameter $w_{\rm eff}$ depends just on $\beta$
and can take any value from $-1$ to $1$, reflecting the relative preponderance
of the potential and kinetic energies of the field $\phi$ on the total energy
density of the Universe. More specifically, for $\beta=0$ the potential term
$V_a e^{-\beta\kappa\phi}$ is preponderant and the scalar field $\phi$ behaves
like a cosmological constant ($w_{\rm eff}=-1$), giving rise to a period of
accelerated expansion of the Universe; for $\beta=\pm\sqrt6$, it is the kinetic
term $\dot{\phi}^2/2$ that dominates and the scalar field behaves as a
stiff-matter fluid ($w_{\rm eff}=1$). Here, the most interesting situation is
the one in which critical point D corresponds to a state of accelerated
expansion ($w_{\rm eff}<-1/3$), occurring for $|\beta|<\sqrt{2}$.

At critical point E, which is always an attractor, the evolution of the
Universe can be dominated by either $\phi$ or $\xi$, depending on the values
of $\alpha$ and $\beta$. For
$\beta=(\sqrt{\alpha^2+24}-\alpha)/2$ [lower boundary of the region of
existence of critical point E, see panel (e) of Fig.~\ref{Fig:critical
points}], the evolution of the Universe is dominated by the potential term of
the scalar field $\phi$ as $\alpha$ tends to infinity ($\Omega_\phi=1$,
$w_{\rm eff}=-1$). For $\beta=\sqrt{\alpha^2+6}$ (upper boundary), it is the
scalar field $\xi$ that dominates the evolution for $\alpha\rightarrow+\infty$
($\Omega_\xi=1$, $w_{\rm eff}=0$). As one approaches the point of intersection
of these two curves, at $(0,\sqrt6)$, the evolution becomes dominated by the
kinetic term of $\phi$ ($\Omega_\phi=1$, $w_{\rm eff}=1$). Here, again, the
most interesting case is the one for which critical point E corresponds to a
state of accelerated expansion ($w_{\rm eff}<-1/3$), occurring for values of
$\alpha$ and $\beta$ lying in the region of the parameter space defined by
the conditions $\beta<\alpha/2$ and $\beta>(\sqrt{\alpha^2+24}-\alpha)/2$ [see
panel (f) of Fig.~\ref{Fig:critical points}]. Once this critical point is
reached, the ratio between the density parameters of the scalar fields is
locked at a constant value which depends on $\alpha$ and $\beta$, namely,
\begin{equation}
\frac{\Omega_\phi}{\Omega_\xi}=
\frac{12+\alpha^2-\beta^2}{2(\alpha\beta+\beta^2-6)}.
 \label{E-ratioDEDM}
\end{equation}
This is the so-called scaling solution, first discussed in
Ref.~\cite{wetterich-1988} and often used to try to solve the coincidence
problem.
Note that, for certain values of the parameters $\alpha$ and $\beta$, more
specifically, for $\beta<\alpha/2$ and
$\beta>(2\sqrt{\alpha^2+18}-\alpha)/3$ ($\alpha>4\sqrt2$),
the critical point E corresponds to an accelerated solution ($w_{\rm eff}<-1/3$) and,
simultaneously, the dynamics of the Universe is dominated by the scalar field
$\xi$ ($\Omega_\xi>\Omega_\phi$). Therefore, in this case, the scalar field
$\xi$ behaves as a dark-energy component, driving accelerated expansion
together with the scalar field $\phi$. However, as already mentioned [see
discussion after Eq.~(\ref{Q-interaction})], the condition $|\alpha-\beta|\lesssim 1$
must be satisfied in order to avoid conflict with primordial nucleosynthesis,
implying that the relevant solutions for the critical point E correspond
to values of $\alpha$ and $\beta$ for which the scalar field $\xi$ does not
behave as dark energy.

Having identified the physical nature of the critical points of the dynamical
system~(\ref{DS}), let us now proceed to the description of the phase
portraits shown in Fig.~\ref{Fig:phase portraits}, which are representative of
each of the nine regions of the parameter space.

In region I of the parameter space $(\alpha,\beta)$, the dynamical
system~(\ref{DS}) has only two critical points, A and B, which are an
attractor and a repeller, respectively. All orbits start on B and end on A,
some of them passing through the zone of the phase space in which the Universe
expansion is accelerated [see panel (a) of Fig.~\ref{Fig:phase portraits}].
Therefore, the initial and final states of the Universe are of stiff-matter
domination with a possible intermediate stage of accelerated expansion.

In region II, all orbits originate at critical point B and end at critical
point C (except for the orbits connecting the B to A and A to C through the
boundaries of the phase space), some of them passing in the acceleration zone
and near the saddle point A [see panel (b) of Fig.~\ref{Fig:phase portraits}].
Therefore, the Universe goes from a stiff-matter initial state---sometimes
through an intermediate stage of accelerated expansion---to a final state
dominated by a scalar field which, depending on the values of $\alpha$ and
$\beta$, has a behavior ranging from dust to stiff matter.

In region III, the attractor is now critical point E, while point B is a
repeller and both A and C are saddle points. Some of the orbits approach C
(which may correspond to a matter-dominated state for certain values of
$\alpha$ and $\beta$), before heading to the attractor E [see panel (c) of
Fig.~\ref{Fig:phase portraits}]. Although critical point E is always located
outside the acceleration zone, some orbits approaching this point may pass
through this zone, giving rise to a temporary period of accelerated expansion.

In region IV, the phase space has no repelling critical points; for
$\tau\rightarrow -\infty$, all orbits approach the heteroclinic one connecting
critical points A and B through the boundaries of the phase space. An orbit
originating, for instance, near critical point A, heads to the vicinity of
point B, before spiraling to the attractor E [see panel (d) of
Fig.~\ref{Fig:phase portraits}], which may correspond to an accelerated
solution if $2\beta<\alpha$.

Region V is the only region of the parameter space for which the corresponding
phase portrait has five critical points (see Table~\ref{Table:9 regions}).
Critical point E is again the attractor, which corresponds to a scaling
solution. Critical points A and B are repellers, near which the dominant
scalar field $\phi$ behaves as stiff matter. Critical points C and D are both
saddle points. In the vicinity of C, the scalar field $\xi$ may be
preponderant, giving rise to a matter-dominated era. In what concerns critical
point D, it may correspond to a state of (temporary) accelerated expansion,
with $-1/2<w_{\rm eff}<-1/3$, if $\beta<\sqrt2$. Some orbits, originating at
points A or B, first approach C and then head to E through the accelerating
zone [see panel (e) of Fig.~\ref{Fig:phase portraits}]. If critical point E
lies inside this zone, accelerated expansion is the final state of the
Universe; otherwise, accelerated expansion is just a temporary stage, before
the Universe evolves to a final state in which $-1/3<w_{\rm eff}<1$.

In region VI, critical point E is again the attractor, which always
corresponds to an accelerated scaling solution. The repelling critical point
is A, while B and D are saddles. Orbits originating near point A, first
approach B, before heading to E [see panel (f) of Fig.~\ref{Fig:phase
portraits}]. Note that critical point C does not exist in this region of the
parameter space; therefore, the Universe cannot experience an intermediate
era of matter domination.

In region VII, the attractor is critical point D, which corresponds to a state
of everlasting accelerated expansion for $|\beta|<\sqrt2$. Point C is a
saddle, near which the dynamics may be dominated by the scalar field $\xi$.
All orbits originating in repelling points A and B (except for the
heteroclinic orbits connecting points A and B to C) end up at critical point D
[see panel (g) of Fig.~\ref{Fig:phase portraits}]. Of particular interest are
those orbits that, before heading to D, closely approach point C. In this
case, the final state of accelerated expansion may be proceeded by a long
period of dark-matter domination.

In region VIII, the final stage of evolution of the Universe corresponds to
critical point D, which, for $\beta>-\sqrt2$, is again an accelerated
solution. The repeller is critical point A, from which all orbits originate
(except for the heteroclinic orbit connecting points B and D through the
boundary of the phase space). Some of the orbits leaving critical point A
first approach B, before heading to D [see panel (h) of Fig.~\ref{Fig:phase
portraits}].

Finally, in region IX, the situation is similar to that of region I, but with
critical points A and B reversing their roles [see panel (i) of
Fig.~\ref{Fig:phase portraits}].

At this point a comment is in order. For clarity of presentation and easier
interpretation of the roles played by each of the two scalar fields in the
cosmic evolution, we have chosen not to overload the equations of motion with
radiation and ordinary baryonic matter. In particular, the inclusion of these
two components would increase the number of equations of the dynamical system
describing our cosmological model, and, consequently, would make its analysis
and interpretation technically more demanding and conceptually less clear. But
since the Universe has had a radiation-dominated era in the past and contains,
in addition to dark matter, ordinary baryonic matter, we should, when further
analyzing the phase portraits corresponding to the different regions of the
parameter space, focus our attention on the later phases of the cosmic
evolution and ignore the early phases, which, as we have seen, correspond to a
stiff-matter state (reflecting the fact that the repellers are always the
critical points A and B, for which $w_{\rm eff}=1$). Therefore, in what
follows, we assume that the Universe has undergone an inflationary period,
followed by a long enough radiation-dominated era encompassing the
nucleosynthesis period, and that the subsequent evolution is described by the
later stages of our two-scalar-field cosmological model, in which the scalar
field $\xi$ accounts for all (dark and baryonic) matter content (yielding
$\Omega_{\xi,0}\approx0.31$) and the scalar field $\phi$ accounts for dark
energy (with $\Omega_{\phi,0}\approx0.69$).

For our model to reproduce the main features of the evolution of the observed
Universe, we should search for solutions of the dynamical system~(\ref{DS})
having an intermediate matter-dominated era, long enough to allow for
structure formation, followed by a present era of (everlasting or temporary)
accelerated expansion.
Such solutions must have the critical point C as a saddle point (the only one
capable of reproducing a long enough intermediate era of matter domination)
and the critical point D or E as a final attractor (the only ones allowing
for accelerated expansion).

Let us start by noting that, at critical point C, the density parameter of the
scalar field $\xi$ is given by
\begin{equation}
 \Omega_\xi=1-\frac{(\alpha-\beta)^2}{6},
\end{equation}
implying that this critical point corresponds to a state of matter domination
only if the difference between the parameters $\alpha$ and $\beta$ is small
enough. In Ref.~\cite{sa-2020b} it has been shown, resorting to numerical
simulations, that $\alpha$ and $\beta$ should satisfy the condition
$|\alpha-\beta|\lesssim 1$ to ensure that the transition between the
radiation- and matter-dominated eras does not occur too early in the cosmic
history, thus avoiding a conflict with primordial nucleosynthesis. In what
follows, we adopt this upper limit, which implies $\Omega_\xi\gtrsim 5/6$ and
$w_{\rm eff}\lesssim1/6$ at critical point C. This choice restricts the
relevant solutions to regions III, V, and VII of the parameter space, the only
ones that overlap, even if only partially, with the strip defined by
$|\alpha-\beta|\lesssim 1$.

In these three regions of the parameter space, for an appropriate choice of
$\alpha$ and $\beta$, the matter-dominated era may be followed by a period of
accelerated expansion.

In region III, the attractor E is always located outside the acceleration
zone (defined by the condition $\beta<\alpha/2$, see Table~\ref{Table:properties CP}),
implying that accelerated expansion can only be temporary (corresponding
to phase-space orbits that, before heading to point E, pass through the
acceleration zone, see Fig.~\ref{Fig:phase portraits}).
These temporary accelerated solutions only exist for points
of the parameter space $(\alpha,\beta)$ sufficiently close to the line
$\beta=\alpha/2$. This condition, together with the condition
$|\alpha-\beta|\lesssim 1$, severely restricts the values of $\alpha$ and
$\beta$ for which the Universe experiences a (temporary) period of accelerated
expansion.

In region V, the attractor E could be chosen to lie inside the acceleration
zone, but the corresponding values of $\alpha$ and $\beta$ would then not
satisfy the condition $|\alpha-\beta|\lesssim 1$; therefore, this attractor
should lie outside the acceleration zone and, again, the state of accelerated
expansion can only be temporary. However, in this case, the restriction to the
allowed values of $\alpha$ and $\beta$ is not as strong as in region III.

Finally, in region VII, the attractor  D corresponds to a state of everlasting
accelerated expansion for $|\beta|<\sqrt2$. This condition, conjugated with
$|\alpha-\beta|\lesssim 1$, allows for a wide choice of values of $\alpha$ and
$\beta$ giving rise to cosmological solutions with the required features. In
this region of the parameter space, temporary accelerated solutions can also
be found for values of $\beta$ slightly above $\sqrt2$ (attractor D slightly
outside the acceleration zone).

\begin{figure*}[t]
  \includegraphics{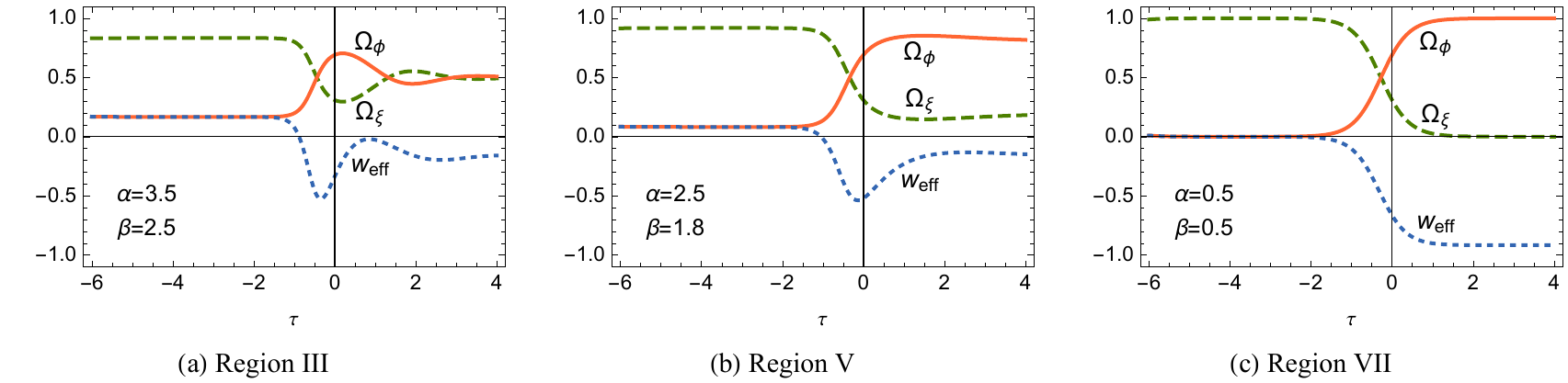}
\caption{\label{Fig:OmegaWeff} Evolution of $\Omega_\phi$, $\Omega_\xi$, and
$w_{\rm eff}$ for values of $\alpha$ and $\beta$ belonging to regions III, V,
and VII of the parameter space. At the present time, $\tau=0$, the density
parameters are $\Omega_{\phi,0}\approx0.69$ and $\Omega_{\xi,0}\approx0.31$ in
all cases, while the effective equation-of-state parameter is $w_{{\rm
eff},0}\approx-0.34$, $-0.52$, and $-0.66$ for the cases shown in panels (a),
(b), and (c), respectively. In regions III and V, the attractor is critical
point E, implying that, for $\tau\rightarrow+\infty$, the ratio between the
density parameters approaches a constant value [$\Omega_\phi/\Omega_\xi=1$ and
$4.3$ for the cases shown in panels (a) and (b), respectively]. In region VII,
where the attractor is critical point D, the energy density of the dark-matter
field $\xi$ quickly approaches zero for $\tau>0$ and, consequently, the
evolution of the Universe becomes entirely dominated by the dark-energy field
$\phi$.}
\end{figure*}

In summary, among the multitude of cosmological solutions of our
two-scalar-field model, those reproducing the main features of the Universe's
evolution lie in regions III, V, and VII of the parameter space. There, for an
appropriate choice of $\alpha$ and $\beta$, the scalar field $\xi$, which
behaves as cold dark matter, dominates the evolution of the Universe for
enough time to allow for structure formation; this stage of evolution is then
followed by an era of accelerated expansion---temporary or permanent---driven
by the scalar field $\phi$, which, therefore, behaves like dark energy.

The evolution of the density parameters $\Omega_\phi$ and $\Omega_\xi$, as
well as the evolution of the effective equation-of-state parameter $w_{\rm
eff}$, for different values of $\alpha$ and $\beta$ belonging to regions III,
V, and VII of the space parameter, is shown in Fig.~\ref{Fig:OmegaWeff}. In
all cases, we choose initial conditions guaranteeing that the transition from
matter to dark-energy domination occurs in a recent past, namely, for
$\tau\approx-0.5$, and also that, at the present time $\tau=0$, the value of
the density parameter of the scalar field $\phi$ is in agreement with
cosmological measurements \cite{Planck-parameters-2015}, namely,
$\Omega_{\phi,0}\approx0.69$. In regions III and V, because the attractor is
critical point E, the ratio $\Omega_\phi/\Omega_\xi$ tends, in the future, to
a constant value, that depends on $\alpha$ and $\beta$. On the contrary, in
region VII, where the attractor is critical point D, the energy density of the
dark-matter field rapidly approaches zero, independently of the values of
$\alpha$ and $\beta$; consequently, in the future, the Universe becomes
entirely dominated by the dark-energy field. In all the cases considered in
Fig.~\ref{Fig:OmegaWeff}, the effective equation-of-state parameter at present
is smaller than $-1/3$, signaling an accelerated growth of the scale factor of
the Universe. In regions III and V, this accelerated expansion is only
temporary; in the future, $w_{\rm eff}$ approaches the quantity
$(\beta-\alpha)/(\alpha +\beta)$, which, for the chosen values of $\alpha$ and
$\beta$, is always greater than $-1/3$. On the contrary, in region VII, the
asymptotic value of the effective equation-of-state parameter depends only on
$\beta$, namely, $w_{\rm}=-1+\beta^2/3$, implying that, for $|\beta|<\sqrt2$,
the accelerated expansion of the Universe lasts forever.

\section{Conclusions\label{Sect-conclusions}}

In this article, we have investigated the late-time evolution of the Universe
within a cosmological model in which dark matter and dark energy are
identified with two interacting scalar fields. More specifically, we assume
that one of the scalar fields, $\xi$, oscillates rapidly around the minimum of
its quadratic potential, thus behaving like a dark-matter fluid, while the
other scalar field, $\phi$, evolving under an exponential potential, gives
rise to the current era of accelerated cosmic expansion, thus behaving like
dark energy.

As shown in Ref.~\cite{sa-2020b}, this two-scalar-field cosmological model
admits viable scenarios for the evolution of the Universe. More specifically,
upon certain assumptions and an appropriate choice of the parameters $\alpha$
and $\beta$, it is possible to obtain a correct sequence of eras in the
evolution of the Universe, namely, an inflationary era driven by the scalar
field $\xi$, a radiation-dominated era encompassing the primordial
nucleosynthesis period, an era dominated by the dark-matter field $\xi$ and,
to a lesser extent, by ordinary baryonic matter, long enough to allow for
structure formation, and, finally, a current era of accelerated expansion
driven by the dark-energy field $\phi$.

Our previous investigations of this two-scalar-field cosmological model
\cite{sa-2020a,sa-2020b} resorted to numerical methods to solve the evolution
equations and, consequently, have not covered the entire parameter space. To
fill this gap, we have now turned to the powerful methods of qualitative
analysis of dynamical systems, applied with great success to astrophysical and
cosmological problems for several decades. With such methods, it is possible
to cover the entire parameter space $(\alpha,\beta)$ and, therefore, ensure
that all cosmological solutions of interest are identified.

Because of the symmetries of action~(\ref{action 2SF}), the parameter space
can be restricted, without any loss of generality, to either the half-plane
$\beta\geq0$ or the half-plane $\alpha\geq0$. We have chosen the latter
possibility and, in addition, opted not to consider the case $\alpha=0$,
corresponding to an action with a standard kinetic term for the scalar field
$\xi$, since this case has been extensively studied in the literature.
Therefore, in our dynamical-system analysis, the parameters $\alpha$ and
$\beta$ span the open half-plane $\alpha>0$.

In this work, we have chosen not to overload the evolution
equations~(\ref{dot-rhoxi})--(\ref{friedmann}) with radiation and ordinary
baryonic matter, in order to better highlight the roles played by the scalar
fields $\xi$ and $\phi$ as dark matter and dark energy. Having made this
choice, we focused our attention on the later phases of cosmic evolution---the
eras dominated by dark matter and dark energy---, assuming that previously the
Universe has undergone an inflationary period and a radiation-dominated era.

The two-dimensional dynamical system~(\ref{DS}), arising from the evolution
equations~(\ref{dot-rhoxi})--(\ref{friedmann}), admits five critical points,
whose stability properties were investigated within the linear theory and,
when this was not feasible, resorting to the center manifold theory and
Lyapunov's method. This stability analysis, carried out for all possible
values of $\alpha$ and $\beta$, has shown that the parameter space is
naturally divided into nine regions (see Fig.~\ref{Fig:9 regions} and
Table~\ref{Table:9 regions}), each of which corresponds to a qualitatively
different phase portrait of the dynamical system.

We have shown that viable solutions---in the sense that they correspond to a
cosmic evolution in which a long enough matter-dominated era is followed by a
current era of accelerated expansion---can be found in regions III, V, and VII
of the parameter space.

Two distinct possibilities have been identified. First, in region VII, the
Universe, after a matter-dominated era, evolves to a state of everlasting
accelerated expansion, in which the energy density of the dark-matter field
rapidly approaches zero and, consequently, the evolution becomes entirely
dominated by the dark-energy field [see panel (c) of
Fig.~\ref{Fig:OmegaWeff}]. Second, in regions III and V, the stage of
accelerated expansion following the matter-dominated era is always temporary
and, for $\tau\rightarrow+\infty$, the ratio between the energy densities of
dark energy and dark matter, given by Eq.~(\ref{E-ratioDEDM}), tends to a
nonzero value [see panels (a) and (b) of Fig.~\ref{Fig:OmegaWeff}].

In both cases, the values of $\alpha$ and $\beta$ should satisfy the condition
$|\alpha-\beta|\lesssim 1$ to ensure that the transition between the
radiation- and matter-dominated eras does not occur too early in the cosmic
history, thus avoiding a conflict with primordial nucleosynthesis
\cite{sa-2020b}. In region VII, this condition, conjugated with the condition
$|\beta|<\sqrt2$ for the existence of a final state of accelerated expansion,
allows for a wide choice of values of the parameters $\alpha$ and $\beta$
giving rise to solutions with the required features. In regions III and V, on
the contrary, this condition is more restrictive, limiting to a small set the
allowed values of $\alpha$ and $\beta$.

The two-scalar-field cosmological model given by action~(\ref{action 2SF}),
arising in a great variety of theories of gravity, like the Jordan-Brans-Dicke
theory, Kaluza-Klein theories, $f(R)$-gravity, string theories, and hybrid
metric-Palatini theories, seems to be quite promising. It allows for a unified
description of inflation, dark energy, and dark matter, which is able to
reproduce, at least qualitatively, the main features of the evolution of the
observed Universe. The results obtained so far within this cosmological model
constitute a first step that needs to be taken further; we expect to do so in
future publications.

\appendix
\section{Stability \protect\\ of the critical points\label{Sect-appendix_1}}

For most values of $\alpha$ and $\beta$, linear theory suffices to assess the
stability of the critical points of the dynamical system~(\ref{DS}). However,
for certain values of these parameters, this is not enough and one has to
resort to other methods, such as the center manifold theory and Lyapunov's
method.  For more details on the methods used in this Appendix, the reader is
referred to the specialized literature
\cite{carr-1982,guckenheimer-1983,bogoyavlensky-1985} and to a recent review
on dynamical systems applied to cosmology \cite{bahamonde-2018}.

\subsection{Critical point A}

At critical point A, with coordinates $x=1$ and $y=0$, the eigenvalues of the
Jacobian matrix of the dynamical system~(\ref{DS}) are
\begin{equation}
 \lambda_1=3+\frac{\sqrt6}{2}(\alpha-\beta)
 \quad {\rm and} \quad
 \lambda_2=3-\frac{\sqrt6}{2}\beta,
\end{equation}
implying that for $\beta>\alpha+\sqrt6$,
$\sqrt6<\beta<\alpha+\sqrt6$, and $\beta<\sqrt6$ the critical point is an
attractor, a saddle, and a repeller, respectively [see panel (a)
of Fig.~\ref{Fig:critical points} and Table~\ref{Table:9 regions}].

For $\beta=\alpha+\sqrt6$ and $\beta=\sqrt6$ one of the eigenvalues vanishes,
forcing us to go beyond the linear theory.

Let us start with the case $\beta=\alpha+\sqrt6$, using Lyapunov's method
to study the stability of the critical point. Consider the function
$V(x,y)=(x-1)^2+y^2$, defined in the phase space of the dynamical
system~(\ref{DS}), i.e., in the upper half of the unit circle centered at the
origin ($x^2+y^2\leq1, y\geq0$). This function is equal to zero at critical
point A and positive elsewhere. Furthermore, its derivative,
\begin{equation}
 V^\prime = -3(1+x)(1-x)^3 -\sqrt6 \alpha y^2 -3y^4,
\end{equation}
is negative in the neighborhood of A (recall that the parameter space is the
open half-plane $\alpha>0$). Therefore, we arrive at the conclusion that the
critical point is asymptotically stable and the phase-space orbits near it are
similar to the ones shown in panel (a) of Fig.~\ref{Fig:phase portraits}.

Let us now turn to the case $\beta=\sqrt6$. Since the eigenvalue
$\lambda_1=\sqrt6\alpha/2$ is positive, the critical point cannot be an
attractor. It is either a saddle or a repeller, depending on the behavior of
the dynamical system on the center manifold, which we now proceed to
investigate. In new variables $u=x-1$ and $v=y$, which shift the critical
point to the origin, the dynamical system~(\ref{DS}) becomes
\begin{subequations}
 \label{DS-a}
  \begin{align}
  u^\prime &= \lambda_1 u + f(u,v),
\\
  v^\prime &= g(u,v),
  \end{align}
\end{subequations}
where the nonlinear functions $f$ and $g$ are given by
\begin{align}
 f(u,v) &= \bigg( 3+\frac{\sqrt6}{4}\alpha \bigg) u^2
         + \frac{\sqrt6}{4}\alpha v^2 +\frac32 u^3 -\frac32 uv^2,\\
 g(u,v) &= \frac32 u^2 v -\frac32 v^3.
\end{align}

The center manifold, obtained as a Taylor series expansion from the equation
\begin{equation}
\frac{dh(v)}{dv} g[h(v),v]-\lambda_1 h(v) - f[h(v),v]=0,
\end{equation}
is given by
\begin{equation}
 u=h(v)=-\frac12v^2-\frac18 v^4 + \mathcal{O}(v^6)
\end{equation}
and the flow on it is governed by the differential equation
\begin{equation}
 v^\prime = \frac32 v \left[ h^2(v)-v^2 \right]
  = -\frac32 v^3 + \mathcal{O} (v^5).
\end{equation}
Therefore, along the $v$ direction the orbits approach the origin, implying
that critical point A is a saddle for $\beta=\sqrt6$ (recall that along the
$u$ direction the orbits move away from critical point A). In the neighborhood
of this point, the orbits are similar to the ones shown in panels (b), (c),
and (d) of Fig.~\ref{Fig:phase portraits}.

In summary, critical point A is an attractor for $\beta\geq\alpha+\sqrt6$, a
saddle for $\sqrt6\leq\beta<\alpha+\sqrt6$, and a repeller for $\beta<\sqrt6$.

\subsection{Critical point B}

At critical point B, with coordinates $x=-1$ and $y=0$, the eigenvalues of the
Jacobian matrix of the dynamical system~(\ref{DS}) are
\begin{equation}
 \lambda_1=3+\frac{\sqrt6}{2}\beta
 \quad {\rm and} \quad
 \lambda_2=3-\frac{\sqrt6}{2}(\alpha-\beta),
\end{equation}
implying that for $\beta>\alpha-\sqrt6$,
$-\sqrt6<\beta<\alpha-\sqrt6$, and $\beta<-\sqrt6$ the critical point is
a repeller, a saddle, and an attractor, respectively
[see panel (b) of Fig.~\ref{Fig:critical points} and Table~\ref{Table:9 regions}].

For $\beta=\alpha-\sqrt6$, one of the eigenvalues, $\lambda_2$, is zero and
the other, $\lambda_1=\sqrt6\alpha/2$, is positive, implying that the
critical point cannot be an attractor. Let us use again the center manifold
theory to determine the behavior of the orbits of the dynamical system in the
vicinity of the critical point. In new variables $u=x+1$ and $v=y$, which
shift the critical point to the origin, the dynamical system~(\ref{DS})
becomes
\begin{subequations}
 \label{DS-b}
  \begin{align}
  u^\prime &= f(u,v),
\\
  v^\prime &= \lambda_1 v + g(u,v),
  \end{align}
\end{subequations}
where the nonlinear functions $f$ and $g$ are given by
\begin{align}
 f(u,v) &= -3 u^2 +\frac{\sqrt6}{2} \alpha v^2
           + \frac32 u^3 - \frac32 u v^2,\\
 g(u,v) &= -\frac{\sqrt6}{2} \alpha u v + \frac32 u^2 v -\frac32 v^3.
\end{align}
For this system, the center manifold, determined from the equation
\begin{equation}
 \frac{dh(u)}{du} f[u,h(u)]-\lambda_1 h(u) - g[u,h(u)]=0
\end{equation} is $v=h(u)=0$.
The flow in this manifold is governed by the equation
\begin{equation}
 u^\prime=-3u^2 +\frac32 u^3,
\end{equation}
from which one concludes that the critical point is a saddle node, i.e., along
the $u$ direction the orbits approach critical point B for positive $u$ and
move away from it for negative $u$. However, one should take into account that
in coordinates $u$ and $v$ the phase space lies entirely on the half-plane
$u\geq0$; therefore, along the $u$ direction all physically relevant orbits
approach the critical point, while along the $v$ direction all orbits move
away from it, meaning that, from the physical point of view, the critical
point can be considered a saddle. In the neighborhood of critical point B, the
orbits are similar to the ones shown in panels (d), (f), and (h) of
Fig.~\ref{Fig:phase portraits}.

For $\beta=-\sqrt{6}$, one of the eigenvalues, $\lambda_1$, is zero and the
other, $\lambda_2=-\sqrt6\alpha/2$, is negative, implying that critical
point B can be either an attractor or a saddle. Let us use Lyapunov's
method to show that it is an attractor. Consider the function
$V(x,y)=(x+1)^2+y^2$ defined on the phase space of the dynamical
system~(\ref{DS}). This function is equal to zero at critical point B and
positive elsewhere. Furthermore, its derivative,
\begin{align}
 V^\prime & = -3(1-x)(1+x)^3-3y^4 \nonumber \\
 &-\frac{\sqrt6}{2} \alpha (1+x)(1-x^2-y^2),
\end{align}
is negative in the neighborhood of B.
Therefore, we arrive at the conclusion that the critical point is
asymptotically stable and the orbits in its vicinity are similar to the
ones shown in panel (i) of Fig.~\ref{Fig:phase portraits}.

In summary, critical point B is a repeller for $\beta>\alpha-\sqrt6$, a saddle
for $-\sqrt6<\beta\leq\alpha-\sqrt6$, and an attractor for $\beta \leq
-\sqrt6$.

\subsection{Critical point C}

Critical point C, with coordinates $x=(\beta-\alpha)/\sqrt6$ and $y=0$, exists
for $\alpha-\sqrt6\leq\beta\leq\alpha+\sqrt6$. The corresponding eigenvalues,
\begin{equation}
 \lambda_1 = \frac{\alpha^2 -\beta^2 +6}{4}
 \quad {\rm and} \quad
 \lambda_2=\frac{(\alpha -\beta)^2 -6 }{4},
\end{equation}
imply that for $\sqrt{\alpha^2+6}<\beta<\alpha+\sqrt6$ and
$\alpha-\sqrt6<\beta<\sqrt{\alpha^2+6}$ critical point C is an attractor and a
saddle, respectively [see panel (c) of Fig.~\ref{Fig:critical points} and
Table~\ref{Table:9 regions}].

For $\beta=\alpha+\sqrt6$, critical point C coincides with A and, therefore,
the conclusions drawn above apply here, i.e., the critical point is
asymptotically stable, attracting all orbits of the phase space, similarly to
the situation depicted in panel (a) of Fig.~\ref{Fig:phase portraits}.

For $\beta=\sqrt{\alpha^2+6}$, the eigenvalue $\lambda_1$ becomes zero and
\begin{equation}
  \lambda_2=-\frac{\alpha}{2} \left( \sqrt{\alpha^2+6}-\alpha\right)
\end{equation}
is negative. To determine whether the critical point is an attractor or a
saddle, let us analyze the flow on the center manifold. In variables $u$ and
$v$, for which the critical point is shifted to the origin, the dynamical
system~(\ref{DS}) is given by
\begin{subequations}
 \label{DS-z}
  \begin{align}
  u^\prime &= \lambda_2 u + f(u,v),
\\
  v^\prime &= g(u,v),
  \end{align}
\end{subequations}
where the nonlinear functions $f$ and $g$ are given by
\begin{align}
 f(u,v) &= \frac{\sqrt6}{2}\big(\sqrt{\alpha^2+6}-\alpha\big) u^2
         +\frac{\sqrt6}{2} \alpha v^2  \nonumber \\
        &+ \frac32 u^3 - \frac32 u v^2,\\
 g(u,v) &= -\frac{\sqrt6}{2} \alpha u v + \frac32 u^2 v -\frac32 v^3.
\end{align}
\phantom{.}
 \vspace{3mm}
\phantom{.}

For this system, the center manifold is approximated by the Taylor series
expansion
\begin{equation}
 u=h(v) = \frac{\sqrt6}{\sqrt{\alpha^2+6}-\alpha} v^2
      + \mathcal{O}(v^4)
\end{equation}
and the flow on it is governed by the equation
\begin{equation}
 v^\prime=-\frac32\left( 1+\frac{2\alpha}{\sqrt{\alpha^2+6}-\alpha} \right)
  v^3 + \mathcal{O} (v^5).
\end{equation}
Since the quantity in parentheses is always positive, we conclude that
critical point C is an attractor for  $\beta=\sqrt{\alpha^2+6}$. In the
neighborhood of this point, the orbits are similar to the ones shown in panel
(b) of Fig.~\ref{Fig:phase portraits}.

Finally, for $\beta=\alpha-\sqrt6$, critical point C coincides with B and,
therefore, the conclusions drawn above apply here, i.e., the critical point is
a saddle and, in its neighborhood, the orbits are similar to the ones shown in
panels (d), (f), and (h) of Fig.~\ref{Fig:phase portraits}.

In summary, critical point C is an attractor for
$\sqrt{\alpha^2+6}\leq\beta\leq\alpha+\sqrt6$ and a saddle for
$\alpha-\sqrt6\leq\beta<\sqrt{\alpha^2+6}$.

\subsection{Critical point D}

Critical point D, with coordinates $x=\beta/\sqrt6$ and
$y=\sqrt{1-\beta^2/6}$, exists for $-\sqrt6\leq\beta\leq\sqrt6$. The
eigenvalues of the Jacobian matrix of the dynamical system~(\ref{DS}) are
\begin{equation}
 \lambda_1 = -3+\frac{(\alpha+\beta)\beta}{2}
 \quad {\rm and} \quad
 \lambda_2=-3+\frac{\beta^2}{2},
\end{equation}
implying that for $(\sqrt{\alpha^2+24}-\alpha)/2<\beta<\sqrt6$
and $-\sqrt6<\beta<(\sqrt{\alpha^2+24}-\alpha)/2$ the critical point is
a saddle and an attractor, respectively
[see panel (d) of Fig.~\ref{Fig:critical points} and
Table~\ref{Table:9 regions}].

For $\beta=\sqrt6$, critical point D coincides with A and, therefore, the
conclusions drawn above apply here, i.e., the critical point is a saddle and
in the neighborhood of this point the orbits are similar to the ones shown in
panels (b), (c), and (d) of Fig.~\ref{Fig:phase portraits}.

For $\beta=(\sqrt{\alpha^2+24}-\alpha)/2$, the eigenvalue $\lambda_1$
becomes zero and
\begin{equation}
  \lambda_2=-\frac{\alpha}{4} \left( \sqrt{\alpha^2+24}-\alpha \right)
\end{equation}
is negative, implying that critical point D is either an attractor or a
saddle. Let us use here the central manifold theory to analyze the stability
of the critical point. In the variables $u=x-\beta/\sqrt6$ and
$v=y-\sqrt{1-\beta^2/6}$, which shift the critical point to the origin, the
dynamical system~(\ref{DS}) becomes
\begin{subequations}
 \label{DS-2}
\begin{align}
  u^\prime &= \sqrt{\frac32\alpha^2 -\lambda_2^2} \, v + f(u,v),
\\
  v^\prime &=  \lambda_2 v + g(u,v),
\end{align}
\end{subequations}
where the nonlinear functions $f$ and $g$ are given by
\begin{align}
 f(u,v) &= \frac{\sqrt6}{4}\sqrt{\alpha^2+24} \, u^2
          -3\sqrt{1-\frac{2\lambda_2^2}{3\alpha^2}} \, u v \nonumber
\\
        & + \frac{\sqrt6}{4} \alpha v^2
          + \frac32 u^3
          -\frac32 uv^2,
\\
 g(u,v) &= \frac32 \sqrt{1-\frac{2\lambda_2^2}{3\alpha^2}} \; u^2
          -\frac92 \sqrt{1-\frac{2\lambda_2^2}{3\alpha^2}} \; v^2 \nonumber
\\
        & +\frac32 u^2v -\frac32 v^3 .
\end{align}

In order to apply the center manifold theorem,
the differential equation for the variable $u$ should not contain linear terms.
Therefore, we perform a new change of variables,
\begin{equation}
w=u+v\sqrt{\frac{3\alpha^2}{2\lambda_2^2}-1} \quad {\rm and} \quad z=v,
\end{equation}
that brings the dynamical system~(\ref{DS-2}) to the required form,
namely,
\begin{subequations}
 \label{DS-3}
\begin{align}
  w^\prime &= F(w,z),
\\
  z^\prime &= \lambda_2 z + G(w,z),
\end{align}
\end{subequations}
where the nonlinear functions $F$ and $G$ are given by
\begin{align}
 F(w,z) &= -\frac{3\sqrt6\alpha}{2\lambda_2} w^2
    + \sqrt{\frac{3}{-\lambda_2}} (3\lambda_2-\alpha^2) w z \nonumber
 \\
 & -\frac{\sqrt6 \alpha}{4\lambda_2}(\alpha^2-2\lambda_2)z^2
   +\frac32 w^3
   -\frac{3\alpha}{\sqrt{-2\lambda_2}} w^2 z \nonumber
 \\
 &-\frac{3}{4\lambda_2}(2\lambda_2+\alpha^2) w z^2,
 \\
 G(w,z) &= \frac{\sqrt{-3\lambda_2}}{2} w^2
    -\frac{\sqrt6 \alpha}{2} w z
    +\frac14 \sqrt{\frac{3}{-\lambda_2}} (6\lambda_2+\alpha^2) z^2 \nonumber
 \\
 &+\frac32 w^2 z
  -\frac{3\alpha}{\sqrt{-2\lambda_2}} w z^2
  -\frac{3}{4\lambda_2}(\alpha^2 +2\lambda_2)z^3.
\end{align}

In these new variables, the center manifold is given by
\begin{equation}
  z=h(w)= \sqrt{\frac{3}{
     \alpha \left( \sqrt{\alpha^2+24} - \alpha \right)
  }} w^2
  + \mathcal{O} (w^3)
\end{equation}
and the flow on it is determined by
\begin{equation}
  w^\prime = \frac{6\sqrt6}{\sqrt{\alpha^2+24}-\alpha} w^2 +
  \mathcal{O} (w^3).
\end{equation}

From the above equation, we conclude that along the $w$ direction the orbits
approach critical point D for negative $w$ and move way from it for positive
$w$, meaning that D is a saddle node. However, one should take into account
that in coordinates $w$ and $z$ the phase space is the upper part of an
ellipse lying entirely on the half-plane $w\leq0$; therefore, all physically
relevant orbits approach critical point D, which is then, from this point of
view, an attractor. In the neighborhood of this point, the orbits are similar
to the ones shown in panels (g) and (h) of Fig.~\ref{Fig:phase portraits}.

For $\beta=-\sqrt6$, critical point D coincides with B and, therefore, the
conclusions drawn above apply here, i.e., the critical point is
asymptotically stable, attracting all the orbits of the phase space, similarly
to the situation depicted in panel (i) of Fig.~\ref{Fig:phase portraits}.

In summary, critical point D is a saddle for
$(\sqrt{\alpha^2+24}-\alpha)/2<\beta \leq \sqrt6$ and an attractor for
$-\sqrt6 \leq \beta \leq (\sqrt{\alpha^2+24}-\alpha)/2$.

\subsection{Critical point E}

Critical point E, with coordinates $x=\sqrt6/(\alpha+\beta)$ and
$y=\sqrt{6+\alpha^2-\beta^2}/(\alpha+\beta)$, exists for
$(\sqrt{\alpha^2+24}-\alpha)/2 \leq \beta \leq \sqrt{\alpha^2+6}$. The
eigenvalues of the Jacobian matrix of the dynamical system~(\ref{DS}) are
\begin{equation}
 \lambda_{1,2} = -\frac{3\alpha}{2(\alpha+\beta)}
  \left( 1\pm\sqrt{1+K} \right),
\end{equation}
where
\begin{equation}
 K=\frac{2(\beta^2+\alpha\beta-6)(\beta^2-\alpha^2-6)}{3\alpha^2}.
\end{equation}
For $(\sqrt{\alpha^2+24}-\alpha)/2 < \beta < \sqrt{\alpha^2+6}$, $K$ is
negative and, therefore, the real parts of the eigenvalues $\lambda_{1}$ and
$\lambda_{2}$  are also negative, implying that point E is an attractor [see
panel (e) of Fig.~\ref{Fig:critical points} and Table~\ref{Table:9 regions}].
If $-1\leq K<0$, the attractor is a node; if $K<-1$, it is a spiral. Since the
former condition corresponds to two thin regions of the parameter space
adjacent to the curves $\beta=\sqrt{\alpha^2+6}$ and
$\beta=(\sqrt{\alpha^2+24}-\alpha)/2$, in most cases critical point E is a
spiral attractor, as shown in panels (c), (d), (e), and (f) of
Fig.~\ref{Fig:phase portraits}.

For $\beta=\sqrt{\alpha^2+6}$, critical point E coincides with C and,
therefore, the conclusions drawn above apply here, i.e., the critical point is
an attractor and, in its neighborhood, the orbits are similar to the ones
shown in panel (b) of Fig.~\ref{Fig:phase portraits}.

Finally, for $\beta=(\sqrt{\alpha^2+24}-\alpha)/2$, critical point E coincides
with D and, thus, all physically relevant orbits approach it. In the
neighborhood of this point, the orbits are similar to the ones shown in panels
(g) and (h) of Fig.~\ref{Fig:phase portraits}.

In summary, whenever critical point E exists, it is an attractor.

\end{document}